# Skills Beget Skills: Evidence from Historical School Reforms Targeting Health and Further Education[*]

Volha Lazuka

University of Southern Denmark, Lund University, and IZA

Peter Sandholt Jensen

Linnaeus University

April 7, 2025

### Abstract

This paper investigates the dynamic complementarity between school health and education reforms implemented in Denmark between 1940 and 1965: the introduction of school doctors and the expansion of secondary education. Using a staggered difference-in-differences approach for multiple treatments, we study the reform effects on individuals' outcomes in the ages 55–64. We find that each reform leads to significant improvements in health and education outcomes, including reduced mortality, fewer hospitalizations, and higher educational attainment. The singular impact of each reform is doubled when both reforms are implemented together, resulting in a 9 percent increase in earnings. These findings underscore high societal returns to complementary school investments in the long term.

[*]Lazuka: vola@sam.sdu.dk, Jensen: peter.sandholtjensen@lnu.se. We thank Meltem Daysal, Mircea Trandafir, and Miriam Wüst for their comments. Lazuka gratefully acknowledges funding from European Union HORIZON-2020, Marie Skłodowska-Curie Individual Fellowship (grant No 101025481). The authors bear sole responsibility for the content of this paper.

# 1 Introduction

School spending is not only the largest government expenditure but also a key social investment, as historical school expansion reforms have been found to yield some of the highest rates of return in long run (Hendren and Sprung-Keyser, 2020). Additionally, children's health has become an important target of school spending, with similarly large long-term benefits found from the nationwide school nutrition reforms (Lundborg et al. (2022); Bütikofer et al. (2018)). A natural question that arises is how both educational and health policies in schools impact children's long-term outcomes—do they create additional value when implemented together? Theoretically, this question relates to the hypothesis in economics that children who benefit from early investments are more likely to benefit from later investments because "skills beget skills" (Cunha and Heckman, 2007). To date, this hypothesis has not been tested with respect to the interaction between school health and education. From a policy perspective, reforms' interactions more accurately reflect real-world conditions, as most school reforms do not occur in isolation but often overlap with one another or with other social reforms.

This paper examines whether the introduction of primary school doctors, who provide regular check-ups, health advice, and treatments, is particularly beneficial for children also exposed to the secondary school reform that expanded and liberalized access to further education, in terms of long-term outcomes. A small body of studies has addressed this interaction hypothesis using quasi-experimental strategies: Fischer et al. (2021), Johnson and Jackson (2019), and Gilraine (2017) explore the effects of overlapping education reforms and find sizable complementary effects on long-run economic outcomes. In contrast, Rossin-Slater and Wüst (2020) find that infant health care and preschool programs act as substitutes—early education primarily benefits children who had not previously received health care. Adhvaryu et al. (2024) also find substitution effects, studying exposure to a rainfall in early life and an education reform. The reason for the contrasting findings may arise from the fact that health-to-education studies, due to imperfect compliance with single reforms, estimate reduced-form interaction effects from a small and, plausibly, specific population. In total, not only has the interaction between school health and education reforms not been thoroughly studied, but there is also currently no evidence that childhood investments in health and education are complementary.

To test for dynamic complementarity between health and other skills, we examine two overlapping yet independent school reforms implemented in Denmark between 1940 and 1965. The first reform targeted students' health by introducing licensed doctors in schools. The second reform aimed at expanding students' access to secondary and further education through the opening of new, exam-free secondary schools. The Danish historical context offers three advantages to our identification strategy: The reduced-form effects of school-based reforms come from perfect compliance, due to primary school education being compulsory. Additionally, the secondary education reform serves as a follow-up to the school doctors' reform, aligning with the hypothesis that later investments should proceed early investments to create complementarities. Finally, both reforms were implemented widely, but not nationwide, and overlapped in timing, though not fully. This allows us to apply a state-of-the-art empirical design to estimate the average treatment effects on the treated municipalities ($ATTs$) using staggered difference-in-differences (DiD) estimators from Callaway and Sant'Anna (2021) and their extension for the case of multiple treatments in Callaway et al. (2024). These new estimators address contamination bias in designs with heterogeneous treatment effects, which generally worsens when interaction effects are included in two-way fixed-effects specifications to study the interaction of reforms (Chaisemartin and D'Haultfœuille, 2023). To our knowledge, our study is the first to explore interaction effects using the latest methodological advancements.

This paper makes four contributions to the existing literature. First, we establish complementary effects of school health and education reforms, supporting the complementarity hypothesis for health and cognitive skills. This evidence broadens the literature on the complementary effects of education programs (Johnson and Jackson (2019); Hinrichs (2010)), including health reforms that are highly effective when combined with education reforms. Second, we contribute to the literature on the long-term effects of school health reforms. While a handful of studies have examined the impact of school meals (Lundborg et al. (2022); Bütikofer et al. (2018)), the im-pact of school doctors—a historically widespread policy—has been explored in only one prior study (Spika, 2023). Third, we examine the effects of a secondary education reform that has received less attention compared to the broader literature on long-term impacts of primary education reforms (Dominguez and Ruffini (2023); Heinesen and Stenholt Lange (2022); Fischer et al. (2022); Mazumder et al. (2021); Fischer et al. (2020); Hyman (2017); Jackson et al. (2016)). Finally, we show that investments in skills in school ages yield long-term benefits similar to preschool investments, often considered especially effective (Lazuka (2023); Wüst (2022); Almond et al. (2018)). We also propose that our study will encourage the use of DiD estimators robust to heterogeneous treatments within a multiple-treatment framework to estimate interaction effects.

## 2 Background

### 2.1 School System in Denmark

We begin the background section by giving a brief overview of the school system at the time of the introduction of school doctors and the school reform.

Figure 1 shows that there were seven years of compulsory schooling at the time of the policies considered (Gjerløff and Jacobsen, 2014). Most children attended school between the ages of 7 and 14. There were differences between rural and urban areas. In rural areas, children would attend school every second day, and classes were not separated by gender. In urban areas, by contrast, children attended school every day, and classes were separated by gender. After the fifth grade, some students qualified for the so-called middle school (introduced in 1903). The middle school (*mellemskolen* in Danish) lasted four years, and students would qualify for either one year of intermediate school or three years of high school. In international terms, middle school is better described as low secondary schooling, so we will use the term "secondary school" instead. The low secondary school was abolished by a school reform in 1958; see Arendt (2005).

## 2.2 Introduction of School Doctors

School doctors were introduced in many countries starting in the 19th century. In Denmark, school doctors were first introduced in Copenhagen in 1897 (Gjerløff and Jacobsen, 2014), and were then gradually rolled out across the country, with 1958 marking the last year a school doctor was hired in a municipality. Healthcare provided by school doctors remained the primary source of health services for school-aged children until 1973, when the public healthcare system became universal (Birk et al., 2024).

The different sources used to date the introduction of a school doctor in a municipality suggest that the examinations conducted by school doctors led to the discovery of tuberculosis in children (which was not treatable until the mid-1940s). Reports from individual school doctors also reveal that some pupils had poor vision and hearing. Gjerløff and Jacobsen (2014) describes the school doctor in Esbjerg as an emblematic example of the work carried out by school doctors: The doctor, assisted by nurses, would assess the child's health and nutrition through physical examinations, including checking blood for hemoglobin levels, urine for sugar and protein, inspecting teeth for cavities, measuring height and weight, and examining the heart, lungs, vision, and hearing. As another example of the school doctor's role, in 1925, school doctors in Copenhagen found that 52 percent of the examined children showed signs of illness, 14 percent were malnourished, and 4 percent were colorblind.

The school doctor was responsible for addressing poor sanitation and inadequate indoor environments, monitoring maltreatment and outbreaks of contagious diseases, and guiding parents. As undernourished children remained prevalent in Danish schools, doctors were involved in the introduction of school meals (adopted in a few cities) and in advising parents on better nutrition for their children. Additionally, the doctor had a medical authority role in identifying and referring children for special education due to physical or psychological disabilities, as well as teaching students about the intimate aspects of health education.

An example of how doctors interacted with families can be seen in their efforts against lice. For some children and mothers, lice were a source of great shame. For others, they were a daily reality, and the doctors' approach depended on their assessment of the family. In the 1940s, girls in Copenhagen were examined for lice four times a year. The procedure was as follows: "... in cases where it is deemed that the child has been infected accidentally, only a printed notification is sent to the home (…) in most such cases, the mothers will immediately begin treating the entire family. In cases where children are regularly infected, no notification is sent home, but they are calmly treated the same day."

Figure 2 provides a visual description of the roll-out of school doctors over time. Notably, there are spikes in the cumulative share of municipalities that introduced school doctors in the 1940s. This coincides with the 1946 law that made school doctors compulsory nationwide. However, despite the law, the employment of school doctors continued to be gradually introduced across municipalities and became a universal standard only in the 1970s.

## 2.3 The 1937 School Law

During the period of the introduction of school doctors, a school reform was passed in 1937, with the main goal of creating a school that promotes and develops children's talents and abilities, strengthens their character, and provides them with useful knowledge (Kongeriget Danmark, 1937). This law introduced the possibility of an exam-free middle school (grades 6 to 9), which we will refer to as exam-free secondary school; see Figure 1. Exam classes had existed before but faced a challenge: a considerable number of students did not take the final exam and dropped out along the way, especially in the third year when compulsory education had ended. The 1937 law emphasized that pupils should be taught the practical aspects of the curriculum and the exam-free secondary school. Gjerløff and Jacobsen (2014) describes the exam-free secondary school as an alternative for pupils who were not strong in theoretical and reading-intensive subjects (i.e., academic orientation) but still wanted to continue learning after seven years of mandatory schooling (i.e., professional orientation).

The secondary school gradually became so popular that more than two-fifths of a cohort in urban schools opted for this path instead of the undivided 7-year school. In Copenhagen, for instance, there were almost as many students in the free secondary school as in the academic exam secondary school at the start of the period. Over time, only about two-fifths of 14-year-old students in urban areas remained in the seven mandatory school years. In rural areas, the numbers were less impressive but followed a similar trajectory.

The 1937 reform also introduced the possibility of an eighth school year if parents represented 15 pupils between the ages of 14 and 18 years and if the local school committee approved; see Figure 1. However, the 8th year was rarely implemented. Minimum teaching hours were also standardized between urban and rural schools. Additionally, the reform likely interacted with other changes to the school system, as it required the establishment of special needs classes before 1948 where possible.

Figure 3 gives a measure of the roll-out of the school reform using an indicator of the building of the exam-free secondary school. The reform likely kept more pupils in school for a longer time period, thereby increasing their exposure to school doctors.

## 2.4 Mechanisms

Our reading of the historical narrative suggests a number of potential mechanisms through which school doctors and the secondary school reform could have affected long-run outcomes.

First, school doctors could have directly impacted the health of pupils, and in this way, their investment in human capital, which would, in turn, affect both their health and earnings. This could have occurred through mortality effects (e.g., prevention of tuberculosis) or morbidity effects (e.g., the use of glasses). Indirectly, healthier students could have contributed to a better learning environment.

Second, the 1937 school reform could have directly affected human capital accumulation for pupils who were not strong enough to attend regular secondary school but instead enrolled in the exam-free secondary school, which focused on preparing them for specific occupations. The school reform also introduced special needs classes, which could have supported some students. Greater human capital accumulation would likely lead to higher earnings and better health in the long run.

Third, there could also be interactions between the two reforms. Students would likely stay longer in school due to the school reform, which would expose them to school doctors for a longer period, potentially affecting their long-term health. It is also possible that school doctors played a role in sorting students for secondary school by providing input on their cognitive abilities, for example.

.

# 3 Data

## 3.1 Register Microdata

The microdata are sourced from Danish administrative registers covering total population in the period from 1980 to 2023, linked through personal identifiers. These registers provide demographic and socioeconomic information for individuals, as well as for their parents and children. The register data on both an individual's birth year and birth municipality are used for linking to the treatment reform data. We exclude migrants and residents of Greenland, as they likely followed different development paths.

Our focus is on the outcomes that measure human capital performance. Therefore, we select earnings, completed schooling (years and degrees), hospitalization status, participation in a disability scheme, and mortality status. Given that the treatment components are well-defined, we align with the epidemiological literature to further categorize deaths and hospitalizations into those caused by infectious and respiratory diseases (targeted by lung check-ups and treatments), circulatory diseases (related to child maltreatment), and eye diseases. To prevent compositional biases from differential mortality and migration effects between the treated and control groups, we treat deceased and out-migrated cases as null values for earnings and as unit values for hospitalizations and early retirement. Since the registers provide data at an annual frequency or higher, we constructed outcomes that are balanced for ages 55 to 64 and then averaged them.

As we are interested in individuals during their working ages, we focus on the age group 55–64 (born between 1925 and 1956). For these cohorts, students started school at age 7 and finished at age 14 (Gjerløff and Jacobsen (2014)), which confines our analysis to the calendar year 1940 as the first year of treatment under the school reforms.

## 3.2 Treatment by School Doctors' and Secondary School Reforms

We have collected data on the treatment provided by school reforms from several sources. Medical journals and reports from the Danish Health Authority provided information on the employment of school doctors in municipalities with annual data each from 1897 to 1970: *Skolehygienisk Tidskrift*, *Ugesrift for Læger*, and *Medicinalberetninger*. The journals also included articles on the specific treatments performed by the school doctors. Danish school yearbooks, *Den Danske Skole Aarbog*, from 1935 to 1970 contained inventories of schools that we used to identify the first years when municipalities built or renovate schools with exam-free classes. To ensure accuracy, we confirmed the starting years of both reforms with various statistical sources and through direct correspondence with all Danish municipal archives (252 in total). In cases where sources conflicted on the starting year (in only 4 percent of cases), we chose the earlier year.

Figures 4 and 5 display maps of Denmark, where municipalities are categorized based on whether they were treated before 1940, never treated, or by the starting year of the school doctors' and secondary school reforms. Out of 1,989 municipalities, 175 introduced school doctors, and 234 built new schools for the first time between 1940 and 1965. A smaller portion of municipalities implemented reforms before 1940 (101 for school doctors and 40 for secondary schools); these municipalities are excluded from the analysis, as they are considered always treated. In total, 1,728 municipalities did not introduce the reforms until 1966 or later.

Figure 6 shows municipalities further categorized based on whether they were treated exclusively by the school doctors' reform, the secondary school reform, or both reforms. A total of 92 municipalities introduced only the school doctors' reform, 151 municipalities introduced only the secondary school reform, and 83 municipalities implemented both reforms.

## 3.3 Municipality Controls

Data on the observable characteristics of municipalities prior to the reforms (1930–1940) are sourced from the municipality archive 1925–1945 (*Kommunedata* (2022)). This archive includes information on the demographics, economy, and parliamentary elections. Additionally, we use data on mortality by cause and age group for the year 1943 from the cause-of-death register (*Dødsårsagsregisteret* (2019)).

## 3.4 Estimation Samples

Our estimation sample consists of 1,557,728 individuals born in Denmark between 1925 and 1956. They attended primary school during the implementation of the school reforms between 1940 and 1965 and are observed in Danish registers at the age of 55. Of these, 1,240,424 individuals were never treated by the reforms. Among individuals who were treated, 75,507 (203,418) were treated only (ever) by the school doctors' reform, 113,886 (241,797) only (ever) by the secondary school reform, and 127,911 by both reforms. To estimate models at the level of treatment variation, we collapsed our data at the municipality-by-birth cohort cells with weights totaling the number of individuals in each cell. Figure **??** presents the averages of the key outcomes of interest, separated by municipalities that were either ever or never treated by the school doctors' reform.

# 4 Empirical Strategy

## 4.1 Identification and Estimation of Treatment Effects

We aim at obtaining average treatment effects of the school doctors' reform and of its joint effects with the secondary school reform. We do so using estimators for staggered DiD designs in Callaway and Sant'Anna (2021) and their extension for the case of multiple treatments in Callaway et al. (2024). Their approach first estimates single group- and time-specific average treatment effects in the treated municipalities (ATT), $\widehat{ATT}(g, t)$, using fixed effects estimators for two periods / two groups. Municipalities that were never treated suit as a comparison group. Then the approach aggregates single $ATTs$, weighted by the size of each treatment cohorts, to produce summary treatment effects. Summary $ATTs$ are produced separately for singular and multiple treatments. Finally, the approach provides inference through a simple multiplier bootstrap procedure that accounts for multiple hypothesis testing and serial correlation of municipalities over time.

Let $G_j = g \in \{1926,\ldots, 1951\}$ represent the treatment cohorts, $t \in \{1925,\ldots, 1956\}$, and let $C_j = 1$ for municipalities never treated by either reform. Let $D_i = d \in \{0, 1\}$ represent municipalities treated by different combinations of the reforms (see Table 1): 1) treated by the school doctors' reform (*overall* effect of school doctors); 2) treated by the school doctors' reform and never treated by the secondary school reform (*singular* effect of school doctors); 3) treated by the secondary school reform and never treated by the school doctors' reform (*singular* effect of secondary school reform); 4) treated by both the school doctors' and secondary school reforms (*joint* effect). The treatment values are always binary. Our treatment data could allow for more combinations, but we focus on the school doctors' reform, the main treatment of interest.

An unconditional estimator for $ATT(g, t, d)$, the average effect of $d$ for cohort $g$ at time $t \geq g$, is:

$$\widehat{ATT}^{un}(g, t, d) = \frac{\sum_j \Delta Y_{j g-1, t} \cdot 1\{G_j = g, D_i = d\}}{\sum_j 1\{G_j = g, D_i = d\}} - \frac{\sum_j \Delta Y_{j g-1, t} \cdot 1\{C_j = 1\}}{\sum_j 1\{C_j = 1\}} \quad (1)$$

where $\Delta Y_{j,g-1,t} \equiv Y_{j,t} - Y_{j,g-1}$. $\widehat{ATT}^{un}(g,t,d)$ identifies cohort $g$'s $ATT(g,t,d)$ under the assumption that the average evolution of outcomes from $g-1$ to $t$ for the entire population if all experienced reform $d$ is equal to the evolution of outcomes that group $d$ actually experienced (*strong parallel trends assumption*). As discussed in Roller and Steinberg (2023), the identification of the effects of multiple treatments requires the independence assumption: equality in the average evolution between the potential outcomes when a complementary reform is absent ($d = 1$) and present ($d = 3$). However, this assumption is satisfied under a *strong parallel trends assumption* (Callaway et al. (2024)).

The $\widehat{ATT}^{un}(g,t,d)$ are for each cohort $G_j = g \in \{1926,\ldots,1951\}$ across all time periods $t \in \{1925,\ldots,1956\}$. For our purposes, we aggregate them into two summary effects. The first is the overall effect of treatment participation, $\theta_w^{un}$, calculated as a weighted average of $ATTs$ across groups and time periods, with group sizes as weights. This estimate is similar to the DiD estimate from two-way fixed effects regressions, where the issue of negative weights is ruled out. The second is the set of effects aggregated by time since reform initiation, $\theta_{es}^d(e)$, representing the weighted average of treatment effects $e$ time periods before and after the reform (event-study effects). These event-study summary estimates serve multiple purposes: to identify the reform effects at the time of initiation ($e = 0$), detect pre-treatment trends ($e < 0$), and analyze the dynamics of the effect ($e > 0$). We obtain standard errors that account for the estimation of multiple average treatment effects and are generated by a multiplier bootstrap procedure clustered by municipality.

## 4.2 Estimation of *Complementary* Effects

In addition to estimating the *singular* and *joint* effects of the reforms, we aim to estimate the *complementary* effects and their inference—the added value of the joint reforms. Following Roth et al. (2023), we estimate the triple differences using the Callaway and Sant'Anna (2021) approach. First, we calculate the difference between each $\widehat{ATT}^{un(con)}(g,t,d=4)$ and $\widehat{ATT}^{un(con)}(g,t,d=2)$ and then aggregate the estimates into summary measures described above. Finally, we obtain the errors through a simple multiplier bootstrap procedure. A similar approach to estimating the complementary effects of multiple treatments is suggested by Chaisemartin and D'Haultfœuille (2023).

## 4.3 Addressing Imbalance

In the context of school reforms under study, the unconditional parallel trends assumption may not hold. Table 2 shows significant pretreatment imbalance between municipalities that implemented school reforms and those that did not. Municipalities that implemented school reforms exhibited different trends: they were becoming wealthier, receiving more public aid, and had populations that voted for social democrats who supported redistribution policies.

The imbalance between comparison groups of municipalities before the reforms suggests they may have been on different development paths, making the unconditional parallel trends assumption unlikely to hold. To address this, we use a complementary approach to select a valid control group of never-treated municipalities: applying the Callaway and Sant'Anna (2021)'s estimator and the regression adjustment procedure from Sant'Anna and Zhao (2020). The procedure estimates the conditional expectation of the outcome for never-treated municipalities and averages these predictions using the empirical distribution of covariates among treated municipalities:

$$\widehat{ATT}^{con}(g, t, d) = \frac{1}{N_1} \sum_{i:D_i=1} \left[ \left(Y_{i,t} - Y_{i,g-1}\right) - \hat{E}\left[Y_{i,t} - Y_{i,g-1} \mid D_i = 0, X_i\right]\right] \quad (2)$$

where $\hat{E}\left[Y_{i,t} - Y_{i,g-1} \mid D_i = 0, X_i\right]$ is the estimated conditional expectation function. After we obtain the conditional $ATTs$'s for all groups and time periods, we use the same aggregation and inference testing as we did for the unconditional ones. Altogether, this approach allows us to impose the strong parallel trends assumption only for subgroups of municipalities *with similar $X_i$* (*conditional strong parallel trends assumption*).

# 5 Results

## 5.1 Main Results

### 5.1.1 Health Outcomes

We begin by presenting the estimates for health outcomes. Table 3 and Figure 7 present both unconditional and conditional *ATTs* aggregated into summary and event-study effects. As shown in Column 1 (representing the *overall* effect), the school doctors' reform led to a 1.2 percentage point decrease in the probability of being deceased by age 56, which is a reduction of 28 percent of the pretreatment mean in municipalities that ever implemented the reform. Columns 2 and 3 show the effects of the *singular* effects of the school doctors and secondary school reforms; however, neither of these effects is statistically significant. Column 4 reveals that the *joint* effects of the school reforms result in a 2 percentage point decrease in the probability of being deceased by age 56 (a 43 percent reduction of the pretreatment mean). Unconditional and conditional estimates are consistent. Figure 7 further supports this, showing that both the *overall* and *joint* reform effects on mortality appear as sharp declines in the first and subsequent event years.

In Table 4 and Figure 8, we examine the effects on the probability of inpatient hospitalizations due to all causes as well as specific causes for individuals aged 55–64. As with mortality, the school doctors' reform produced both *overall* effects, amounting to 1.1 percentage points, and *joint* effects, amounting to 2 percentage points. Event-study graphs illustrate the emergence of these effects in the first event year onward. We further present the effects for diagnoses related to treatments provided by school doctors. The beneficial effects are observed for infectious and respiratory causes (a 0.6 percentage point decrease) and circulatory causes (a 0.44 percentage point decrease), suggesting that school doctors' reforms prevented maltreatment and tuberculosis in school ages. For eye diseases, however, the reforms increase the probability of hospitalization. When separating planned and acute visits, we find that these effects are primarily driven by planned visits, suggesting that the reforms encouraged preventive behavior regarding eye problems. Since specific diagnosis effects account for about half of the overall hospitalization effects, we conclude that the reforms also contributed to improved overall health.

### 5.1.2  Schooling Outcomes

In Table 5 and Figure 9, we present the results for the schooling outcomes. The summary $ATTs$ estimates indicate that both the school doctor and secondary school reforms yield a positive *overall* effect on the years of schooling: 0.106 years and 0.193 years, respectively. The event-study effects for secondary school reform show an increasing pattern across event years, suggesting that the secondary school reform was beneficial for younger students. Additionally, the reforms led to a *joint* increase in the years of schooling by 0.149 years.

We further find that the increase in the years of schooling was primarily driven by the redistribution of students from those with only a primary or vocational degree to those with a professional or higher degree, as shown in Table 10 and Figure 10. Skilled labor was in short supply but highly demanded by the Danish industrial sector and other specialized fields after World War II, creating a notable education premium (Kærgård, 2022). These conditions, which remained in place for our cohorts during their youth, made pursuing education increasingly attractive. We find that both the school doctor and secondary school reforms increased the probability of obtaining a professional or higher degree by 1.6 percentage points (17% of the pretreatment mean) and 1.3 percentage points (12% of the pretreatment mean), respectively. The *joint* impact of the reforms resulted in a 2.6 percentage point increase (24% of the pretreatment mean).

### 5.1.3  Labor Market Outcomes

We finally present the effects on labor market outcomes for individuals aged 55–64, the key indicators of human capital performance. In Table 8 and Figure 11, we show the results for labor market exit due to disability. The estimates indicate beneficial effects for all types of school doctor and secondary school reforms. The *singular* effects of the school doctor reform and secondary school reform correspond to a decrease in the probability of being on disability by 2 percentage points (6% of the pretreatment mean). The event-study graphs indicate no significant pretrends. *Jointly*, the reforms yield a 2.3 percentage point decline in disability (7 percent of pretreatment mean).

The school reforms have shifted the whole distribution of real earnings for the treated individuals to the right. Table 7 presents the summary effects estimates for the categories of the real earnings' rank: below the 25th percentile (a decrease of 1.6 percentage points or 7.6% of the pretreatment mean), between the 25th and 75th percentiles (an increase of 6.6 percentage points or 3.6% of the pretreatment mean), and above the 75th percentile (an increase of 2.2 percentage points or 8.4% of the pretreatment mean). As an *overall* effect of the school doctors' reform, individuals are less likely to fall into the lowest group and are more likely to be found in the middle and upper groups. These effects arise from both the *singular* and *complementary* impacts of the school doctors' reform.

Regarding real earnings, we find that the school doctor reform produces both an *overall* and *joint* effect, as shown in Table 9 and Figure 12. Due to the school doctor reform, real earnings for individuals aged 55–64 increased by 11 thousand Danish kroner (7% of the pretreatment mean). *Singular* effects of the reforms are positive but not statistically significant. Due to both the school doctor and secondary school reforms, real earnings increased by 14 thousand Danish kroner (9% of the pretreatment mean).

### 5.1.4 *Complementary* Effects

As a culmination of our empirical approach, in Table 10 we present the estimates for the *complementary* effects of the school reforms—the value added by the secondary school reform to the value of the school doctors' reform across all outcomes studied. The results reveal strong and statistically significant effects for the majority of the outcomes, with robust effects emerging for mortality, hospitalizations, professional degree attainment, and real earnings. The inception of secondary education reform alongside the health reform during the schooling years *additionally* increases real earnings by 6.2–7.2 thousand Danish kroner (or 3.8–4.4 percent of the pre-treatment mean). These effects are large in magnitude, approaching the effects of the single health and education reforms targeting children, which have been considered highly effective in the longer term (Wüst (2022); Hendren and Sprung-Keyser (2020)).

## 5.2 Interaction with Other Overlapping Events and Reforms

### 5.2.1 World War II

Our study cohorts include the war years during which Denmark was under German occupation, from 1940 to 1945. Previous research has not identified any significant disruptive effects on the war cohorts in terms of BMI or height (Pedersen et al. (2023); Angell-Andersen et al. (2004)). However, historical accounts depict a Danish school system that faced challenges during the occupation—such as resource shortages and psychological stress on students, particularly those completing their final years—but nonetheless continued to function (Gjerløff and Jacobsen, 2014). For example, a school psychologist writing in 1946 observed that only a few children appeared to be permanently affected by the occupation, particularly those who had directly witnessed traumatic events.

We further examine the interaction effects between the school doctors' reform and World War II, using the same empirical approach as in Section 4. To capture local variation in wartime disruption, we use data from the 1939 parliamentary elections, which indicate the share of votes cast for the German Minority Party across municipalities (*Kommunedata* 2022). A higher vote share is interpreted as reflecting local resilience and a willingness to cooperate with the occupying forces, which may have led to less disruption during the occupation (Lumans, 1993). We define municipalities that voted for the German Minority Party as ever treated by cooperation with the Nazis, with treatment starting from the cohort born in 1927, who were 14 years old at the start of the occupation. In total, 198 municipalities—approximately 7 percent of all—are classified as ever treated, while the remaining municipalities are considered never treated. The assignment of the school doctors' reform remains unchanged from earlier sections.

As the earnings estimates in Panel A of Table 11 show, both the *overall* ($d^* = 1$) and the *singular* effect of the school doctors' reform ($d^* = 2$) remain unaffected by World War II. The *singular* effects of World War II ($d^* = 3$) are (unexpectedly) negative in the unconditional estimations, but in the conditional estimations become small and statistically imprecise. We therefore conclude that World War II had neither disruptive nor interactive effects on the cohorts in our study.

### 5.2.2 Polio Epidemic

The epidemic of acute poliomyelitis struck Denmark in 1952–1953, affecting 7,616 individuals, mostly children under the age of 15 (Lassen, 1953). As a result of polio infection, most children experienced cold-like fever, but a small proportion (1 in every 200 infections) became paralyzed or died. Treatment for polio remained palliative until the vaccine was developed in the 1960s. No noticeable long-term health or economic effects were observed in the polio survivors, likely due to the relatively small population that suffered permanent damage (Serratos-Sotelo et al., 2019). In Danish schools, children with suspected polio infection were monitored by school doctors for their lung and overall health. As a result, we might expect some remedial effects, although these could be mild due to the lack of effective treatment.

Taking the same approach as in the previous section, we estimated the interaction effects between exposure to polio epidemics and the school doctors' reform. Municipalities exposed to the polio epidemic were those with above the upper quartile of mortality due to acute poliomyelitis in 1952–1953 (*Dødsårsagsregisteret* 2019). We denote the first exposed group of municipalities starting from 1938 (they were 14 years old in 1952) and the remaining municipalities as never exposed to the polio epidemic.

As the estimates in Panel B of Table 11 show, there are no *singular* and *joint* effects of the exposure to polio epidemic on earnings ($d^* = 3$ and $d^* = 4$). The effects of the school doctors' reform remain unchanged compared to the main results.

### 5.2.3  Infant Care Reform

During the study period (1937–1949), some Danish municipalities implemented an infant care reform that provided home visits by trained nurses. These nurses encouraged mothers to breastfeed, maintain a clean home environment, and referred ill infants to doctors. As shown by Hjort et al. (2017) and Wüst (2012), the reform significantly improved infant survival rates and led to positive long-term health outcomes among survivors in their middle ages. However, there were no long-term effects on educational or labor market outcomes. Moreover, the infant care reform predates the school doctors' reform by at least six years—an interval that, according to theory, is too long for meaningful interactive effects between the two interventions to emerge (Heckman, 2000). Nevertheless, we examine potential interaction effects between the school doctors' reform and the infant care reform, using treatment assignments for the latter as defined by Wüst (2012).

The estimates in Panel C of Table 11 show that the *singular* effects of the infant care reform on earnings ($d^* = 3$) are small and statistically imprecise, consistent with the findings of Hjort et al. (2017). The effects of the school doctors' reform remain unaffected by the infant care reform. Consequently, the *joint* effects of the two reforms ($d^* = 4$) are also small and lack stability across the unconditional and conditional estimations.

### 5.2.4  School Meals

From 1902, Danish municipalities were allowed to distribute food prepared in school kitchens to poor students during the winter months. By 1953, 93 municipalities, primarily cities, had established mandatory school meals for all children (Statistiske Departament, 1953). However, school meals never became as widely implemented in Denmark as they were in other Nordic countries, where they are a regular part of schoolchildren's daily routines. Our data on the roll-out of school meals across municipalities (collected from statistical sources and correspondence with local archives) shows that most treated municipalities introduced school meals before our period of interest, with only 16 municipalities introducing school meals between 1925 and

1965.

We present the estimates in Panel D of Table 11, showing the *singular* and *joint* effects of the school doctors' and school meals' reforms. The results indicate that school meals did not have an independent effect on long-term earnings but were beneficial only when combined with the presence of school doctors.

Altogether, our results from estimating the interaction effects of the school doctors' reforms with other significant events and reforms, aside from the secondary school reform, have two key implications. First, war, epidemic, and social reforms do not introduce bias into our estimates of the school reforms' interaction effects presented in earlier sections. Second, health interventions that are separated by a long interval—such as the infant care and school doctors' reforms—do not appear to generate additional long-term benefits when combined.

## 5.3 Robustness

### 5.3.1 Spillovers to Siblings

The $ATTs$ estimates would be biased if the school reforms generated spillovers to older siblings, who were not treated by the reforms but whose younger siblings were treated. In general, we do not expect this to be an issue because our comparison sample consists of never-treated municipalities. However, the issue could arise if a substantial portion of families migrate from never-treated to ever-treated municipalities between different births. Since we have parental identifiers, we find that less than 1 percent of families do so, suggesting that this is not a concern. Nevertheless, we exclude these individuals from the estimation sample and rerun the estimations. As shown in Table 12 Panel A, the estimates excluding older siblings are almost identical to our main estimates.

### 5.3.2   Selection into Adult Sample

Since individuals under study are included in the estimation sample at the age of 55, our estimates may suffer from selective mortality. Generally, studies examining the long-term survival of cohorts exposed to different socioeconomic conditions in childhood conclude that the weakest members of the cohorts tend to die before being observed in the registers, meaning that the bias related to selective mortality is typically downward (Allanson and Petrie, 2021). However, this argument is related to the variation across cohorts. In our DiD estimations, bias due to selective mortality is less likely to arise but could occur if mortality patterns vary systematically across both cohorts and municipalities in a way that coincides with the roll-out of the school reforms.

To account for this potential bias, we adjust our models using the two-stage Heckman correction procedure (Heckman, 1979). More specifically, we take advantage of the fact that all our individuals are observed before the age of 55 in the Housing and Population Census of 1970. In the first stage, we model the probability of being included in the estimation sample as a function of housing and family conditions using a probit model. We then predict an inverse Mills ratio for each individual and include it as a covariate in the estimation model. We also tested an alternative correction strategy—inverse probability weighting—with weights derived from predictions based on the same factors (Bailey et al. (2020), Weuve et al. (2012)). However, our results were nearly identical to those obtained with the Heckman correction. We report our Heckman-adjusted estimates in Table 12 Panel B, which are similar to our main results.

## 6   Conclusions

Focusing on two landmark reforms in Denmark between 1940 and 1965, the introduction of school doctors and the expansion of secondary education, and rich administrative register data between 1970 and 2023, this paper investigates whether these reforms generate complementary benefits in terms of long-term outcomes. The school doctors' reform targeted children's health through regular medical check-ups and treatments, while the secondary school reform aimed to expand access to lower secondary education. The overlap of these reforms presents an ideal setting to test a dynamic complementarity hypothesis: whether early health investments increase the effectiveness of later educational investments. The paper uses a DiD design to estimate the effects of these reforms, with municipal-level data spanning several decades, and applies cutting-edge DiD estimators for the case of multiple treatments robust to heterogeneous treatment effects. We also thoroughly examine the robustness of our estimates.

Our findings reveal significant complementary effects between the two reforms, captured in the individuals' outcomes in the ages 55–64. The introduction of school doctors led to significant improvements in long-term health outcomes, including a reduction in mortality and hospitalization rates, particularly for infectious diseases like tuberculosis and circulatory diseases like brain hemorrhage. The expansion of secondary education resulted in increased years of schooling and a higher likelihood of obtaining professional or higher degrees. The combination of the school doctors' and secondary education reforms yielded strong complementary effects, doubling singular effects of each reform. Regarding earnings, the complementary effect of the reforms amounts to a 4 percent increase, approaching the effects of welfare reforms targeting children, which have been considered highly effective in the long term. The evidence presented in this paper underscores the potential for combined school health and education reforms to create lasting societal benefits.

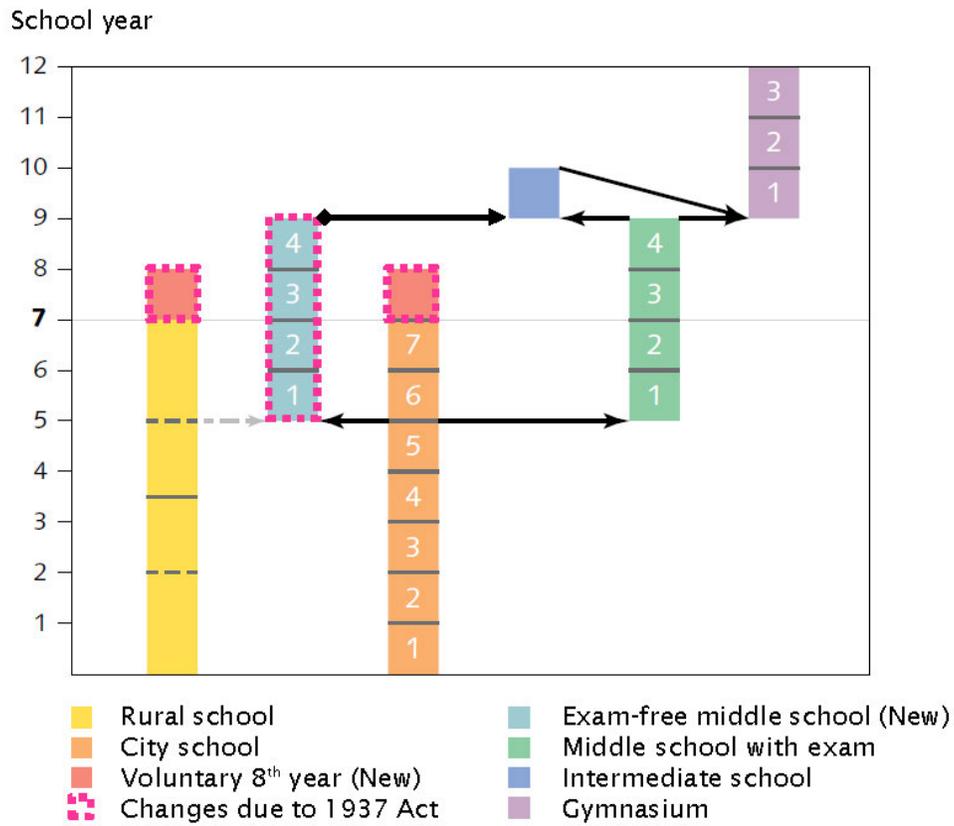

Figure 1: The Danish School System and its Changes due to the 1937 School Act

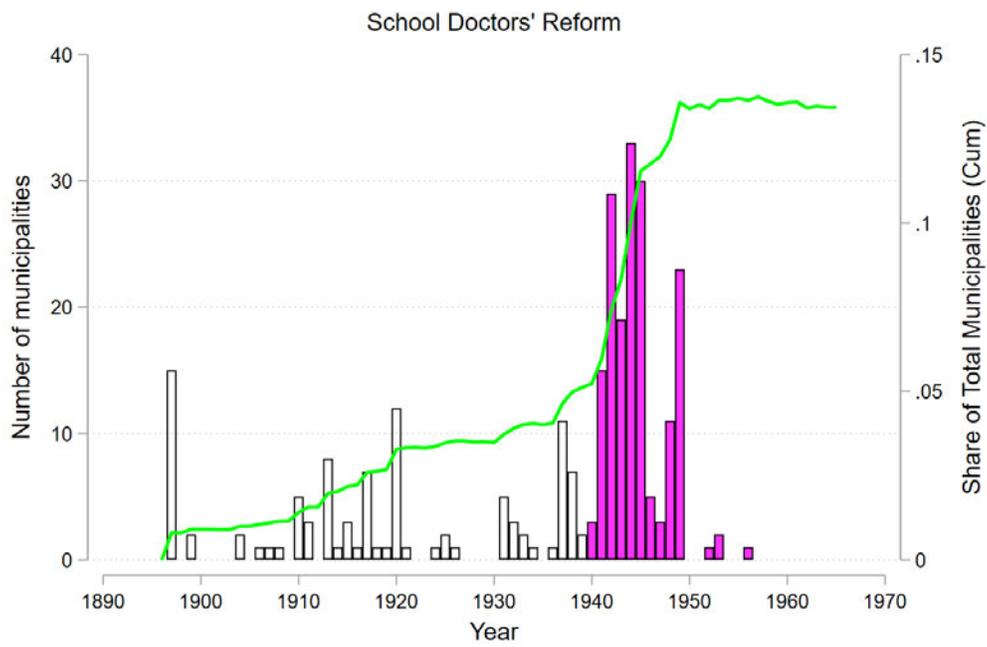

Figure 2: Entry of Municipalities into the School Doctors' Reform

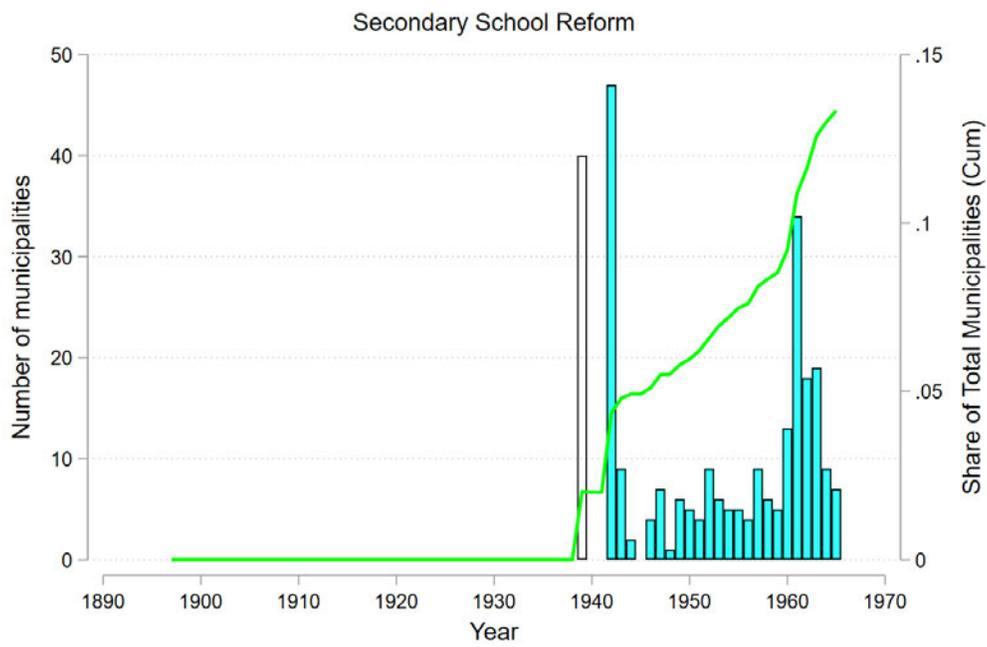

Figure 3: Entry of Municipalities into the Secondary School Reform

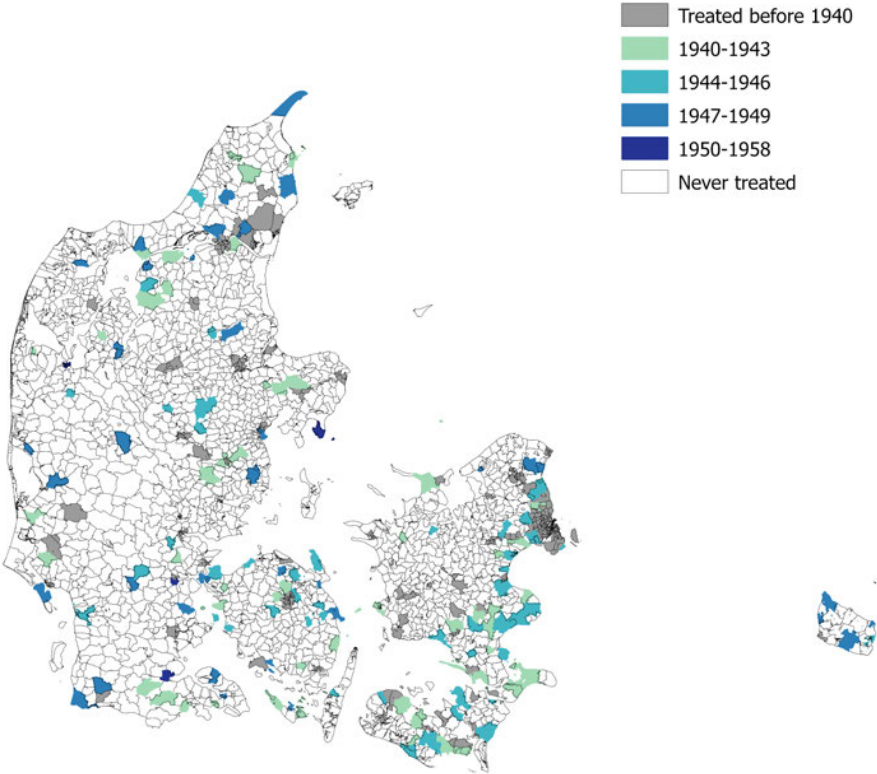

Figure 4: Timing of the School Doctor's Reform

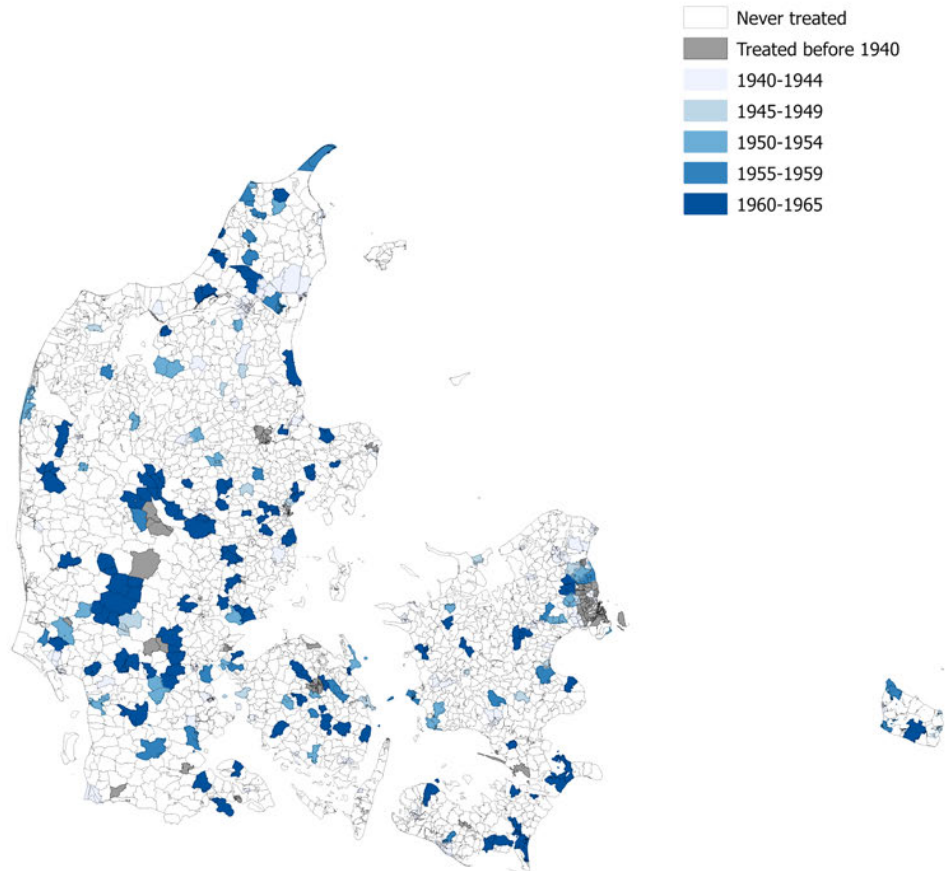

Figure 5: Timing of Secondary School Reform

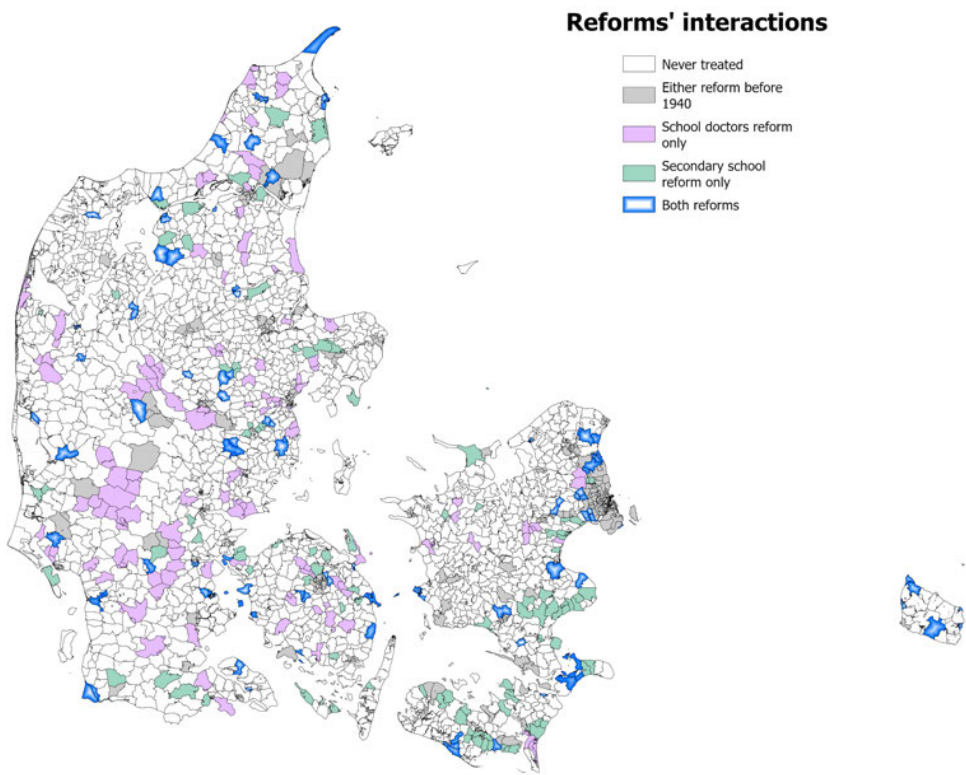

Figure 6: Timing of Reforms' Interactions

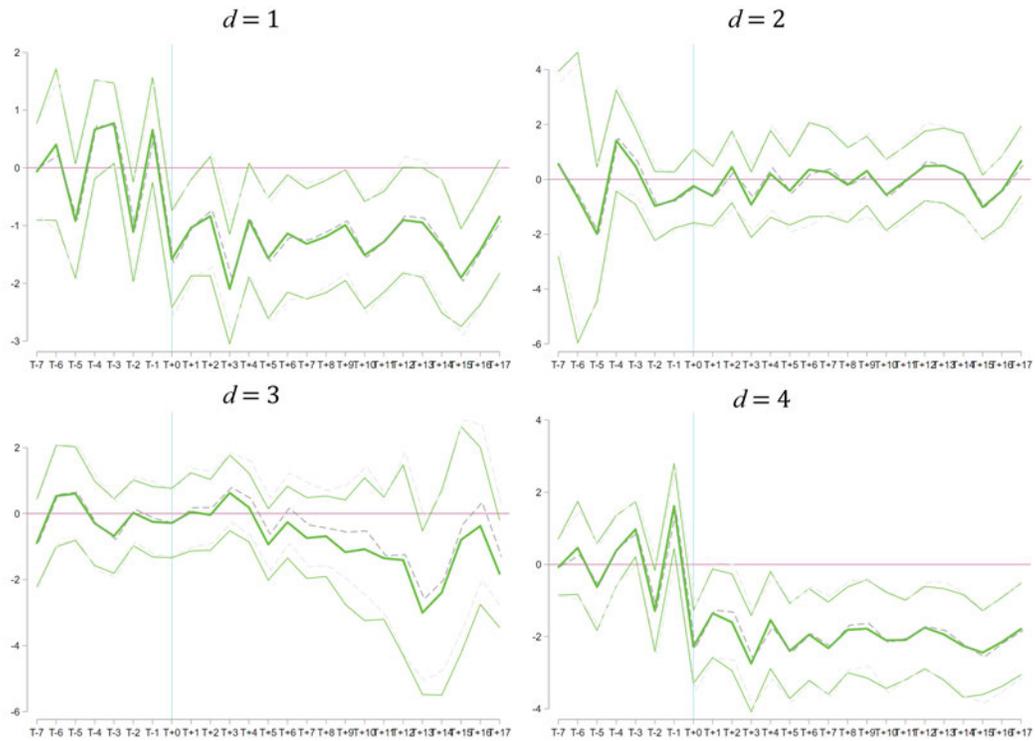

Figure 7: Summary Estimated ATT effects by Event Years for Being Deceased by the Age of 56 ×100: $\theta_{es}^{un}$ (Bold) and $\theta_{es}^{con}$ (Dashed Line)

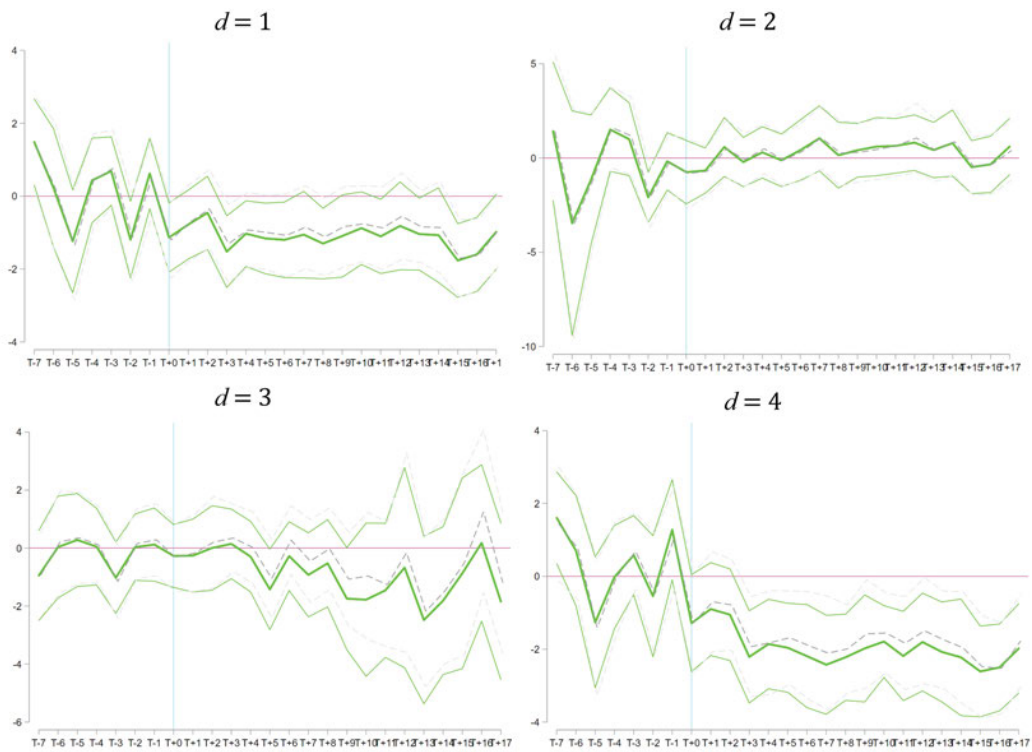

Figure 8: Summary Estimated ATT effects by Event Years for Being Hospitalized in Ages 55–64 ×100: $\theta_{es}^{un}$ (Bold) and $\theta_{es}^{con}$ (Dashed Line)

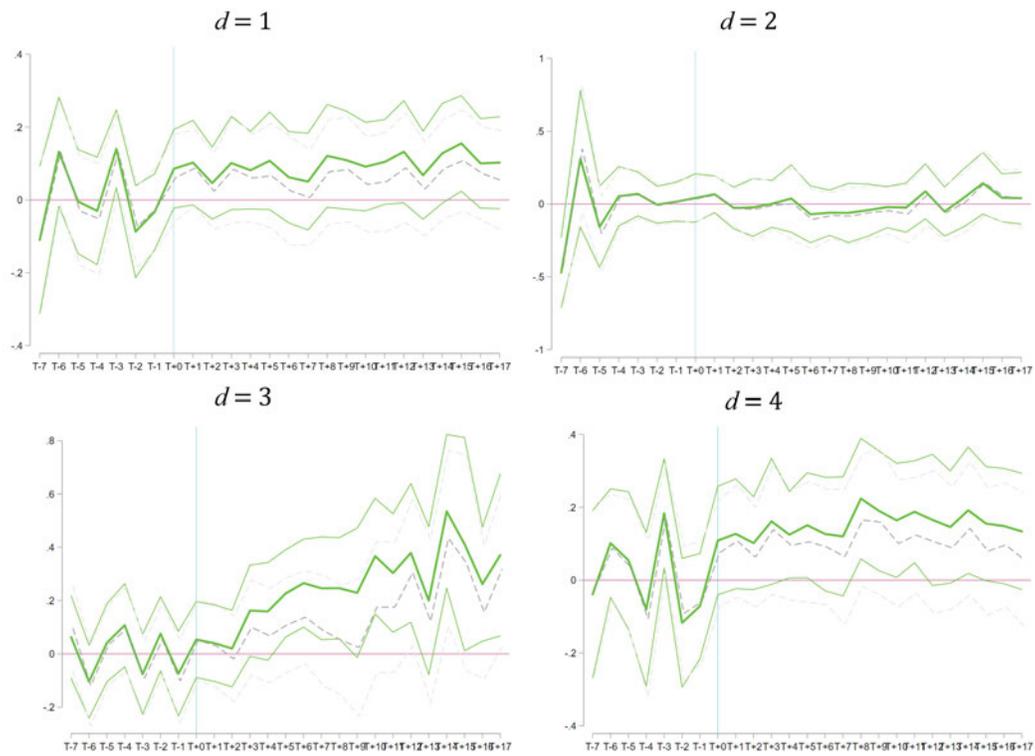

Figure 9: Summary Estimated ATT effects by Event Years for Years of Schooling: $\theta_{es}^{un}$ (Bold) and $\theta_{es}^{con}$ (Dashed Line)

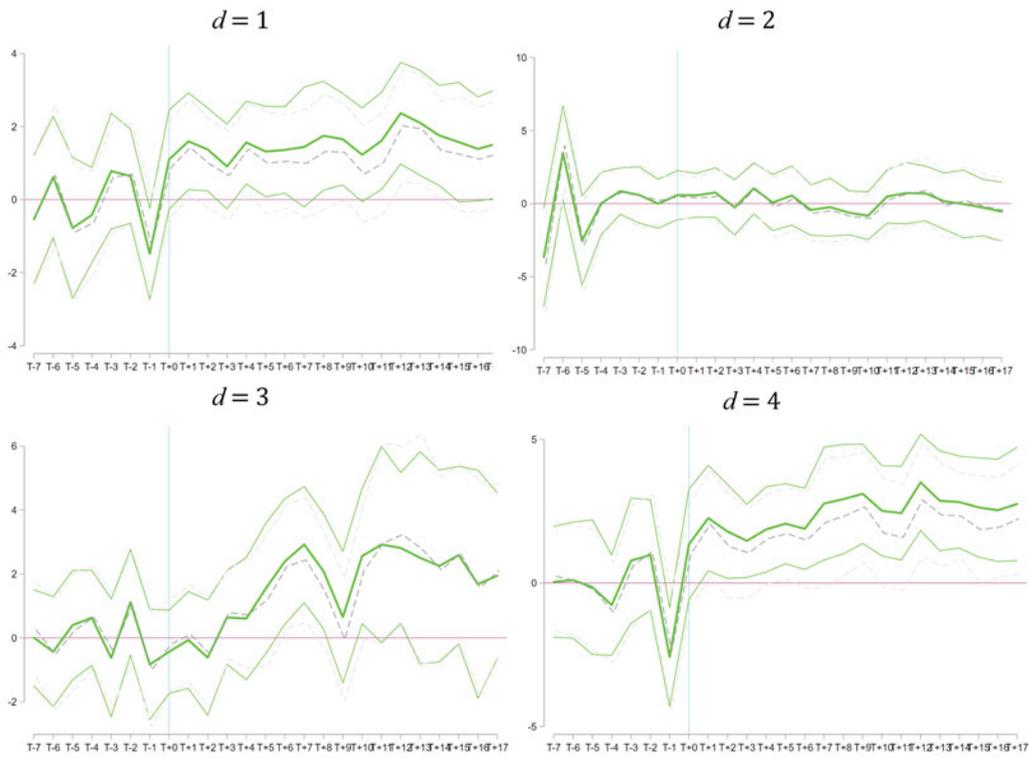

Figure 10: Summary Estimated ATT effects by Event Years Having a Professional or Higher Degree ×100: $\theta_{es}^{un}$ (Bold) and $\theta_{es}^{con}$ (Dashed Line)

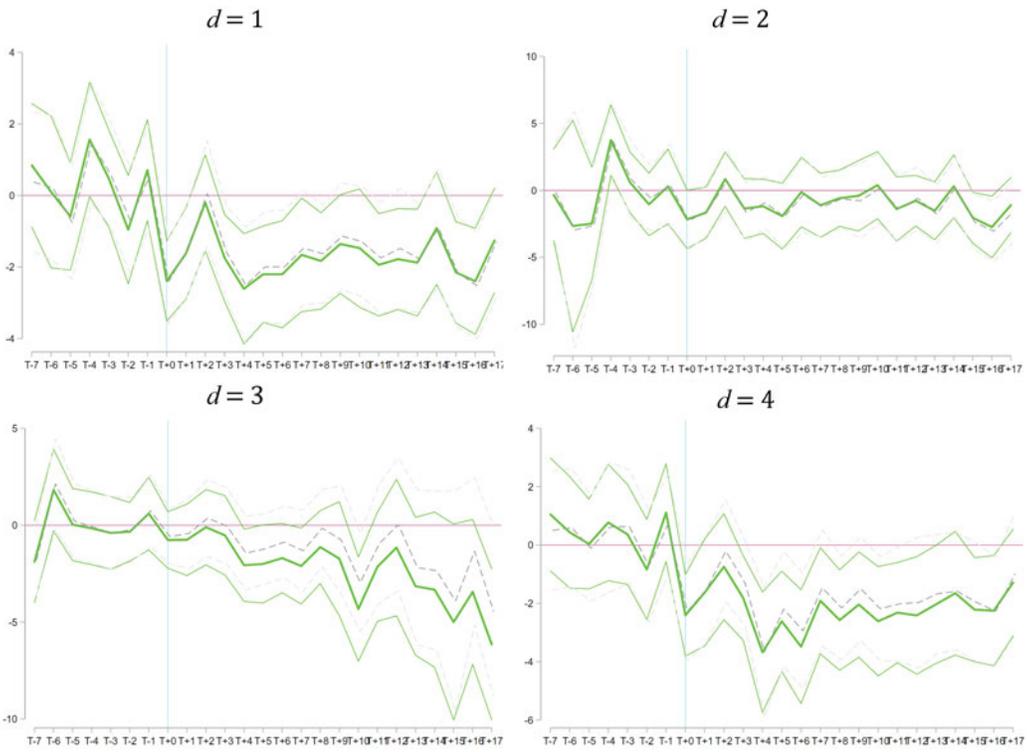

Figure 11: Summary Estimated ATT effects by Event Years for Being in Disability Scheme in Ages 55–64 ×100: $\theta_{es}^{un}$ (Bold) and $\theta_{es}^{con}$ (Dashed Line)

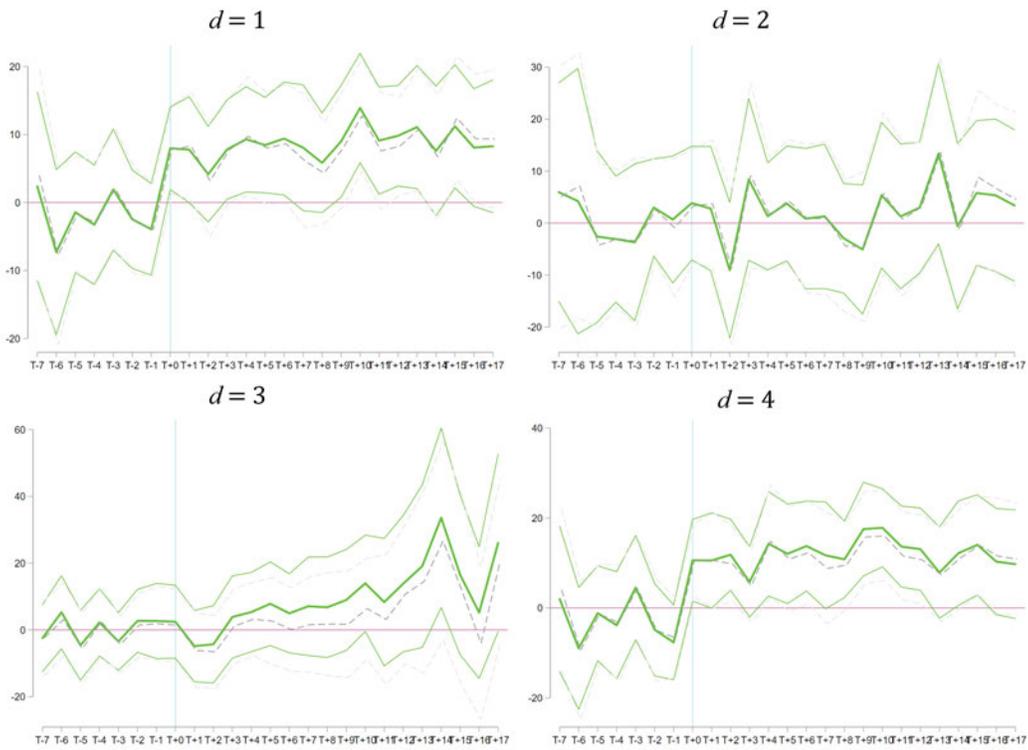

Figure 12: Summary Estimated ATT effects by Event Years for Real Earnings in Ages 55–64 ÷1000: $\theta_{es}^{un}$ (Bold) and $\theta_{es}^{con}$ (Dashed Line)

Table 1: Summary ATT for different types of treatments

|  | $d = 1$ | $d = 2$ | $d = 3$ | $d = 4$ |
| --- | --- | --- | --- | --- |
| Never treated by both reforms | Control | Control | Control | Control |
| Treated by school doctors only | Treatment | Treatment |  |  |
| Treated by secondary school reform only |  |  | Treatment |  |
| Treated by both reforms | Treatment |  |  | Treatment |

Table 2: Balance in Pretreatment Characteristics between Municipalities with and without School Reforms

|  | Difference | $p$-value |
|---|---|---|
| *Levels in 1935* | | |
| Log population | -0.0232 | < 0.10 |
| Log income per capita | -0.0641 | < 0.08 |
| Public aid per capita | 0.885 | < 0.12 |
| Property tax per capita | 2.991* | < 0.01 |
| Share women | 0.657 | < 0.18 |
| Share voting for social-democratic party | 0.352* | < 0.01 |
| Share working in industry | 0.227 | < 0.08 |
| *Changes between 1940 and 1935* | | |
| Log population | 0.133 | < 0.17 |
| Log income per capita | 0.143* | < 0.02 |
| Public aid per capita | 3.291* | < 0.01 |
| Property tax per capita | 4.261* | < 0.02 |
| Share women | 0.454 | < 0.51 |
| Share voting for social-democratic party | 0.603* | < 0.04 |
| Share working in industry | 0.123 | < 0.56 |

Notes: *$p < 0.05$.

Table 3: Summary Estimated ATT Effects of School Reforms on Being Deceased by Age 56 ×100

|  | $d = 1$ (1) | $d = 2$ (2) | $d = 3$ (3) | $d = 4$ (4) |
| --- | --- | --- | --- | --- |
| $\theta_w^{un}$ | -1.215* | -0.087 | -0.668 | -1.995* |
| se | 0.350 | 0.424 | 0.496 | 0.479 |
| $\theta_w^{con}$ | -1.238* | -0.131 | -0.362 | -2.012* |
| se | 0.359 | 0.469 | 0.529 | 0.502 |
| Pre-mean | 4.254 | 3.493 | 5.411 | 4.619 |
| Individuals | 1,443,842 | 1,315,931 | 1,354,310 | 1,368,335 |
| N | 59,849 | 57,713 | 58,088 | 57,021 |

Notes: *$p < 0.05$.

Table 4: Summary Estimated ATT Effects of School Reforms on Being Hospitalized in Ages 55–64 ×100, for All and Specific Diagnoses

|  | $d = 1$ | $d = 2$ | $d = 3$ | $d = 4$ |
|---|---|---|---|---|
|  | (1) | (2) | (3) | (4) |
| *All Causes* | | | | |
| $\theta_w^{un}$ | -1.116* | 0.143 | -0.645 | -2.013* |
| se | 0.381 | 0.452 | 0.534 | 0.526 |
| $\theta_w^{con}$ | -1.038* | 0.083 | -0.259 | -1.834* |
| se | 0.375 | 0.507 | 0.584 | 0.508 |
| Pre-mean | 16.703 | 15.493 | 16.832 | 17.284 |
| *Infectious and Respiratory Diseases* | | | | |
| $\theta_w^{un}$ | -0.319* | 0.002 | -0.292 | -0.547* |
| se | 0.151 | 0.164 | 0.205 | 0.232 |
| $\theta_w^{con}$ | -0.322* | 0.001 | -0.182 | -0.544* |
| se | 0.148 | 0.152 | 0.216 | 0.238 |
| Pre-mean | 16.703 | 15.493 | 16.832 | 17.284 |
| *Circulatory Diseases* | | | | |
| $\theta_w^{un}$ | -0.365* | -0.218 | 0.020 | -0.441* |
| se | 0.114 | 0.155 | 0.138 | 0.157 |
| $\theta_w^{con}$ | -0.387* | -0.274 | 0.003 | -0.445* |
| se | 0.118 | 0.172 | 0.147 | 0.176 |
| Pre-mean | 16.703 | 15.493 | 16.832 | 17.284 |
| *Eye Diseases* | | | | |
| $\theta_w^{un}$ | 0.006 | -0.026 | -0.039 | 0.024* |
| se | 0.021 | 0.060 | 0.053 | 0.007 |
| $\theta_w^{con}$ | 0.014 | -0.019 | -0.040 | 0.035* |
| se | 0.017 | 0.044 | 0.050 | 0.012 |
| Pre-mean | 16.703 | 15.493 | 16.832 | 17.284 |
| Individuals | 1,443,842 | 1,315,931 | 1,354,310 | 1,368,335 |
| N | 59,849 | 57,713 | 58,088 | 57,021 |

Notes: *$p < 0.05$.

Table 5: Summary Estimated ATT Effects of School Reforms on Years of Schooling

|  | $d = 1$ | $d = 2$ | $d = 3$ | $d = 4$ |
|---|---|---|---|---|
|  | (1) | (2) | (3) | (4) |
| $\theta_w^{un}$ | 0.106* | 0.044 | 0.193* | 0.149* |
| se | 0.041 | 0.063 | 0.075 | 0.050 |
| $\theta_w^{con}$ | 0.081 | 0.039 | 0.098 | 0.113* |
| se | 0.044 | 0.069 | 0.078 | 0.058 |
| Pre-mean | 8.382 | 8.010 | 8.576 | 8.562 |
| Individuals | 1,443,842 | 1,315,931 | 1,354,310 | 1,368,335 |
| N | 59,849 | 57,713 | 58,088 | 57,021 |

Notes: *$p < 0.05$.

Table 6: Summary Estimated ATT Effects of School Reforms on Having a Professional and Higher Degree ×100

|  | $d = 1$ (1) | $d = 2$ (2) | $d = 3$ (3) | $d = 4$ (4) |
|---|---|---|---|---|
| $\theta_w^{un}$ | 1.624* | 0.372 | 1.266* | 2.554* |
| se | 0.541 | 0.636 | 0.646 | 0.696 |
| $\theta_w^{con}$ | 1.326* | 0.344 | 1.149 | 2.021* |
| se | 0.529 | 0.666 | 0.640 | 0.716 |
| Pre-mean | 9.413 | 7.132 | 10.876 | 10.508 |
| Individuals | 1,443,842 | 1,315,931 | 1,354,310 | 1,368,335 |
| N | 59,849 | 57,713 | 58,088 | 57,021 |

Notes: *$p < 0.05$.

Table 7: Summary Estimated ATT Effects of School Reforms on Distribution of Real Earnings' Rank in Ages 55–64 ×100

|  | $d = 1$ | $d = 2$ | $d = 3$ | $d = 4$ |
|---|---|---|---|---|
|  | (1) | (2) | (3) | (4) |
| *Below 25th Percentile* | | | | |
| $\theta_w^{un}$ | -1.622* | -0.708 | -1.288 | -2.184* |
| se | 0.543 | 0.775 | 0.756 | 0.746 |
| $\theta_w^{con}$ | -1.634* | -0.745 | -0.385 | -2.218* |
| se | 0.591 | 0.751 | 0.769 | 0.767 |
| Pre-mean | 21.549 | 20.698 | 20.369 | 21.957 |
| *Between 25th and 75th Percentiles* | | | | |
| $\theta_w^{un}$ | 6.508* | 5.847* | 1.957 | 6.814* |
| se | 1.950 | 2.101 | 3.190 | 2.773 |
| $\theta_w^{con}$ | 6.823* | 6.037* | 0.091 | 7.258* |
| se | 2.434 | 2.287 | 3.225 | 3.245 |
| Pre-mean | 177.737 | 171.601 | 172.932 | 180.684 |
| *Above 75th Percentile* | | | | |
| $\theta_w^{un}$ | 2.169* | 1.949* | -0.631 | 2.271* |
| se | 0.650 | 0.948 | 1.071 | 0.924 |
| $\theta_w^{con}$ | 2.274* | 2.012 | -0.001 | 2.419* |
| se | 0.811 | 1.067 | 1.083 | 1.082 |
| Pre-mean | 25.912 | 23.867 | 24.290 | 26.894 |
| Individuals | 1,443,842 | 1,315,931 | 1,354,310 | 1,368,335 |
| N | 59,849 | 57,713 | 58,088 | 57,021 |

Notes: *$p < 0.05$.

Table 8: Summary Estimated ATT Effects of School Reforms on Being on Disability Scheme in Ages 55–64 ×100

|  | $d = 1$ (1) | $d = 2$ (2) | $d = 3$ (3) | $d = 4$ (4) |
|---|---|---|---|---|
| $\theta_w^{un}$ | -1.950* | -1.467* | -2.014* | -2.298* |
| se | 0.551 | 0.732 | 0.944 | 0.700 |
| $\theta_w^{con}$ | -1.932* | -1.688* | -1.192 | -2.101* |
| se | 0.527 | 0.740 | 1.030 | 0.738 |
| Pre-mean | 32.995 | 33.815 | 26.719 | 32.602 |
| Individuals | 1,443,842 | 1,315,931 | 1,354,310 | 1,368,335 |
| N | 59,849 | 57,713 | 58,088 | 57,021 |

Notes: *$p < 0.05$.

Table 9: Summary Estimated ATT Effects of School Reforms on Real Earnings in Ages 55–64 ÷1000

|  | $d = 1$ | $d = 2$ | $d = 3$ | $d = 4$ |
|---|---|---|---|---|
|  | (1) | (2) | (3) | (4) |
| $\theta_w^{un}$ | 10.530* | 6.158 | 6.780 | 13.416* |
| se | 3.096 | 4.942 | 5.603 | 3.907 |
| $\theta_w^{con}$ | 11.021* | 7.399 | 2.385 | 13.517* |
| se | 3.474 | 5.123 | 5.524 | 4.388 |
| Pre-mean | 159.372 | 151.158 | 176.779 | 163.315 |
| Individuals | 1,443,842 | 1,315,931 | 1,354,310 | 1,368,335 |
| N | 59,849 | 57,713 | 58,088 | 57,021 |

Notes: *$p < 0.05$.

Table 10: Summary Estimated *Complementary* ATT Effects of School Reforms on the Outcomes in Ages 55–64

|  | Deceased ×100 | Hospitalized ×100 | Schooling ×100 | Professional ×100 | Disability ×100 | Earnings ÷1000 |
|---|---|---|---|---|---|---|
|  | (1) | (2) | (3) | (4) | (5) | (6) |
| $\theta_w^{un}$ | -1.908* | -2.156* | 0.105* | 2.553* | -0.831* | 7.201* |
| se | 0.691 | 0.532 | 0.042 | 1.087 | 0.309 | 2.871 |
| $\theta_w^{con}$ | -1.881* | -1.873* | 0.074 | 1.025* | -0.413 | 6.118* |
| se | 0.643 | 0.682 | 0.043 | 0.541 | 0.198 | 2.483 |
| Pre-mean | 4.619 | 17.284 | 8.562 | 10.508 | 62.602 | 163.315 |
| Individuals | 1,368,335 | 1,368,335 | 1,368,335 | 1,368,335 | 1,368,335 | 1,368,335 |
| N | 57,021 | 57,021 | 57,021 | 57,021 | 57,021 | 57,021 |

Notes: *$p < 0.05$.

Table 11: Interaction with Overlapping Events and Reforms: Estimated ATT Effects of School Reforms on Real Earnings in Ages 55–64 ÷1000

|  | $d^* = 1$ | $d^* = 2$ | $d^* = 3$ | $d^* = 4$ |
|---|---|---|---|---|
|  | (1) | (2) | (3) | (4) |
| *(A) World War II* | | | | |
| $\theta_w^{un}$ | 11.634* | 10.765* | -8.622* | 5.583 |
| se | 3.703 | 3.593 | 3.777 | 5.554 |
| $\theta_w^{con}$ | 12.937* | 12.549* | -3.585 | 11.876* |
| se | 3.557 | 3.779 | 4.444 | 5.492 |
| Pre-mean | 160.316 | 160.320 | 200.261 | 160.266 |
| Individuals | 1,348,430 | 1,340,714 | 1,232,334 | 1,179,049 |
| N | 55,719 | 55,387 | 55,083 | 51.033 |
| *(B) Polio Epidemic* | | | | |
| $\theta_w^{un}$ | 10.164* | 9.610* | 0.457 | 12.037 |
| se | 3.755 | 3.918 | 5.370 | 6.924 |
| $\theta_w^{con}$ | 11.186* | 11.266* | -10.644 | 11.862 |
| se | 3.225 | 3.474 | 5.403 | 6.789 |
| Pre-mean | 160.316 | 160.320 | 200.261 | 160.266 |
| Individuals | 1,348,430 | 1,340,714 | 1,232,334 | 1,179,049 |
| N | 55,719 | 55,387 | 55,083 | 51.033 |
| *(C) Infant Care Reform* | | | | |
| $\theta_w^{un}$ | 10.353* | 9.173 | -5.972 | 5.583 |
| se | 5.002 | 5.820 | 4.907 | 5.551 |
| $\theta_w^{con}$ | 10.877* | 10.162* | -1.553 | 11.876* |
| se | 3.716 | 4.687 | 6.115 | 5.492 |
| Pre-mean | 160.316 | 157.809 | 221.041 | 162.396 |
| Individuals | 763,278 | 667,289 | 1,262,897 | 682,170 |
| N | 33,569 | 31,005 | 56,084 | 31,115 |
| *(D) School Meals* | | | | |
| $\theta_w^{un}$ | 10.576* | 11.244* | -2.007 | 7.658 |
| se | 3.515 | 3.384 | 4.026 | 10.329 |
| $\theta_w^{con}$ | 11.244* | 11.061* | -0.577 | 14.891 |
| se | 3.537 | 3.189 | 7.347 | 15.338 |
| Pre-mean | 157.906 | 156.797 | 186.227 | 165.647 |
| Individuals | 1,435,494 | 1,409,031 | 1,242,322 | 1,266,887 |
| N | 59,721 | 59,561 | 54,840 | 54,941 |

Notes: *$p < 0.05$.

Table 12: Robustness Analyses for Summary Estimated ATT Effects of School Reforms on Real Earnings in Ages 55–64 ÷1000

|  | $d = 1$ | $d = 2$ | $d = 3$ | $d = 4$ |
|---|---|---|---|---|
|  | (1) | (2) | (3) | (4) |
| *(A) Spillovers to Siblings* | | | | |
| $\theta_w^{un}$ | 10.581* | 6.307 | 6.723 | 13.422* |
| se | 3.083 | 4.989 | 5.863 | 3.934 |
| $\theta_w^{con}$ | 11.088* | 7.574 | 2.424 | 13.498* |
| se | 3.402 | 5.046 | 5.491 | 4.375 |
| Pre-mean | 159.372 | 151.159 | 176.869 | 163.316 |
| Individuals | 1,437,358 | 1,312,666 | 1,349,942 | 1,363,608 |
| N | 59,836 | 57,707 | 58,077 | 57,012 |
| *(B) Heckman Two-Stage Correction Procedure* | | | | |
| $\theta_w^{un}$ | 10.357* | 5.839 | 6.276 | 13.395* |
| se | 2.986 | 4.759 | 5.355 | 3.438 |
| $\theta_w^{con}$ | 10.944* | 7.314 | 3.203 | 13.488* |
| se | 2.995 | 4.724 | 5.355 | 4.202 |
| Pre-mean | 159.372 | 151.158 | 176.779 | 163.315 |
| Individuals | 1,443,842 | 1,315,931 | 1,354,310 | 1,368,335 |
| N | 59,849 | 57,713 | 58,088 | 57,021 |

Notes: *$p < 0.05$.